\documentclass[twocolumn,aps,prd,superscriptaddress,nofootinbib,longbibliography]{revtex4-2}
\usepackage{amsmath,latexsym}
\usepackage{xcolor}
\usepackage[%
  colorlinks=true,
  urlcolor=blue,
  linkcolor=blue,
  citecolor=blue
]{hyperref}
\usepackage{etoolbox}
\usepackage{breqn}

\makeatletter
\let\cat@comma@active\@empty
\makeatother
\usepackage{datetime}% 

\newcommand{\jj}{{\bf j}}
\newcommand{\kk}{{\bf k}}

\newcommand{\rr}{{\bf r}}

\newcommand{\JJ}{{\cal J}}

\newcommand{\be}{\begin{equation}}
\newcommand{\ee}{\end{equation}}
\newcommand{\ba}{\begin{eqnarray}}
\newcommand{\ea}{\end{eqnarray}}
\newcommand{\bse}{\begin{subequations}}
\newcommand{\ese}{\end{subequations}}
\newcommand{\beq}{\begin{eqnarray}}
\newcommand{\eeq}{\end{eqnarray}}
\newcommand{\ds}{\displaystyle}

\newcommand{\sh}{{\bf s}}

\begin{document}
\title{Diffusion of light in turbid media and Kubelka-Munk {theory$^1$}}
\author{Walter Schirmacher}
\affiliation{Institut f\"ur Physik, Staudinger Weg 7, Universit\"at Mainz, D-55099 Mainz, Germany}
\affiliation{Center for Life Nano Science @Sapienza, Istituto Italiano di Tecnologia, 295 Viale Regina Elena, I-00161, Roma, Italy}
\author{Giancarlo Ruocco}
\affiliation{Dipartimento di Fisica, Universita' di Roma ``La Sapienza'', P'le Aldo Moro 5, I-00185, Roma, Italy}
\affiliation{Center for Life Nano Science @Sapienza, Istituto Italiano di Tecnologia, 295 Viale Regina Elena, I-00161, Roma, Italy}
\begin{abstract}
	We show that
	the Kubelka-Munk equations for the description
	of the intensity transfer of light in turbid media
	are equivalent
	to a one-dimensional diffusion equation,
	which is obtained by averaging the three-dimensional
	diffusion equation over the lateral directions.
	This enables us to identify uniquely the Kubelka-Munk
	parameters and derive expressions for diffuse reflection
	and transmission coefficients including the
	effect of internal reflections.
	Without internal reflections we recover the Kubelka-Munk
	formulas for these coefficients.
	We show that the Kubelka-Munk equations are the proper
	radiative-transfer equations for the one-dimensional
	diffusion problem and
	comment on previous attempts to derive the Kubelka-Munk
	equations.
\end{abstract}
\maketitle
%\tableofcontents
\section{Introduction}
\footnotetext{On the occasion of the 60th birthday of Taras Bryk}
Investigating the
reflectance and transmission of turbid media
is a widely-used tool for
materials characterization with applications ranging from soil science,
over
medicine, 
the production of paper and paint, to the design
of laser car headlights
\cite{hecht70,orel95,pasikatan01,torrent08,chang19}. In the analysis
of the observed spectra the theory of diffuse reflectance
and transmissance
of Kubelka and Munk \cite{kubelkamunk31,kubelka48,kubelka54},
has been widely used.
The microscopical
significance of the
phenomenological
parameters $S$ and $K$ appearing in this theory was discussed
in many treatments
\cite{ishimaru78,hecht70,edstroem03,edstroem04,edstroem07,edstroem12,djimbeg11,djimbeg12,elton14,sandoval14},
but with differing results for these coefficients.

Here we show that for a geometry of rectangular incidence onto
a slab, made of turbid material, in which the scattering is strong enough to
lead to diffusive motion of the light intensity, the Kubelka-Munk equations 
are equivalent
to the one-dimensional projection of the 3-dimensional diffusion
equation of the light intensity in
the medium. This is done in the second section.
In the third section we derive expressions for the
diffuse reflectance and transmission coefficients,
including the effect of internal reflection. The standard
Kubelka-Munk results without internal reflection \cite{kubelkamunk31,kubelka48}
are recovered.
In the fourth section we show that the Kubelka-Munk equations
are, in fact, the proper radiative-transfer equations for the
quasi-onedimensional scattering problem 
In the fifth section, we discuss why other
authors might have obtained results for the Kubelka-Munk coefficients different from ours. In the sixth section some conclusions are drawn.

\section{Diffusion and Kubelka-Munk equations}
In the diffusion approximation 
\cite{ishimaru78,porra97} the light intensity $U(\rr)$ 
and the current density $\jj(\rr)$ obey the
steady-state energy-balance and Fick equations
\ba\label{diff1}
\nabla\jj(\rr)&=&
-\lambda_a U(\rr)+\JJ(\rr)\nonumber\\
\nabla U(\rr)
&=&-\frac{1}{\widetilde D}
\jj(\rr)
\ea
which are equivalent to the diffusion equation
\be\label{diff1a}
\lambda_aU(\rr)=\widetilde D
\nabla^2U(\rr)
+\JJ(\rr)
\ee
Here $\JJ(\rr)$ is a source term.

	The quantity $\widetilde D$, which
is the
diffusivity divided by the light velocity in the material\footnote{%
	$c$ is the light velocity and $n$ is the index of refraction.}
$v=c/n$
is given by \cite{diff_durian98}
\be\label{diffu}
\widetilde D=D/v=\frac{1}{\lambda_a+3\lambda_t}
\ee
$\lambda_a,\lambda_s$ and $\lambda_t$ are
the inverse mean free paths due to
absorption, scattering and transport. The latter two
are related as
\be
\lambda_t=\lambda_s(1-\langle \cos\gamma\rangle)
\ee
where $\gamma$ is the scattering angle and
$\langle\cos\gamma\rangle$ is the anisotropy parameter.
 
The relation of the diffusivity to the absorption parameter 
$\lambda_a$, Eq. (\ref{diffu}) had been subject to a dispute in the
literature. It was argued 
\cite{diff_furutsu94,diff_nakai97,diff_durduran97}
that the time-dependent diffusion equation
\be\label{diff1b}
\bigg(v\frac{\partial}{\partial t}+
\lambda_a\bigg)U(\rr,t)=\widetilde D
\nabla^2U(\rr,t)
+\JJ(\rr)\,,
\ee
with a diffusivity that depends on $\lambda_a$,
violates the property, obeyed by the radiative transfer equation,
that the absorptivity $\lambda_a$ should always occur together with
the time derivative in combination
$v\frac{\partial}{\partial t}+
\lambda_a$. Therefore it was argued in Refs. 
\cite{diff_furutsu94,diff_nakai97,diff_durduran97} that the diffusivity
should not depend on the absorptivity $\lambda_a$.
The counter argument is, that the proper
generalization of the steady-state diffusion equation Eq. (\ref{diff1a}) is
{\em not} Eq. (\ref{diff1b}), but a damped telegrapher's equation
\cite{diff_durian98}, which obeys the proper scaling.
However, for this property to be obeyed,
the absorptivity dependence of the diffusivity is given
by (\ref{diffu}) and not by 
$\widetilde D=[3(\lambda_a+\lambda_t)]^{-1}$ according to
the conventional literature (e.g. \cite{ishimaru78}).

\begin{figure}
	\includegraphics[width=0.35\textwidth]{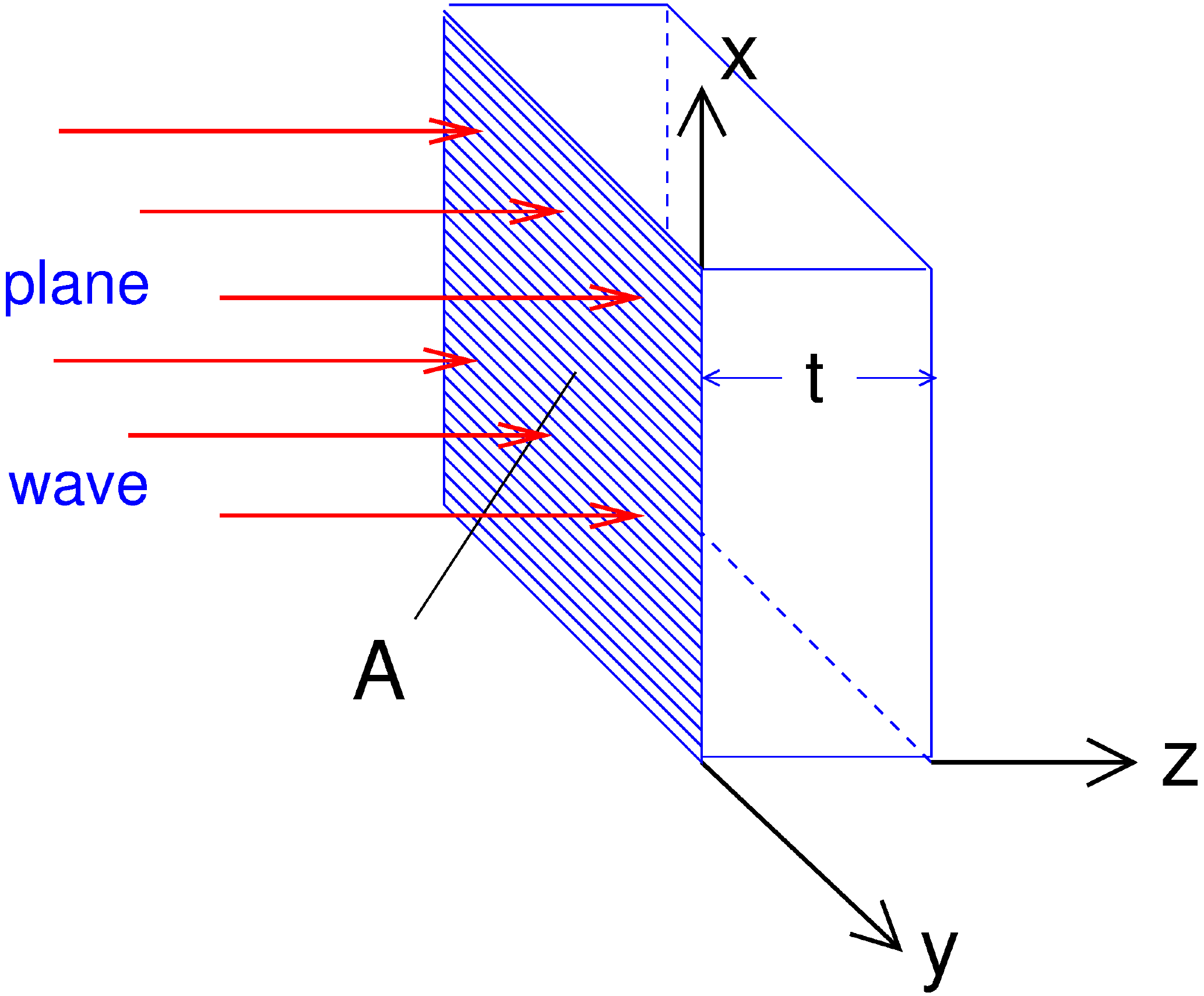}
	\caption{Geometry for the discussion of
	diffuse reflectance and transmission with uniform illumination
	(plane-wave incidence). We consider a slab
	of thickness $t$, which is infinitely extended
	in $x$ and $y$ direction.}
	\label{fig1}
\end{figure}

Let us now consider the geometry of a diffusive-reflection 
(or -transmission) setup with
uniform illumination, i.e.
an incoming plane wave in the $z$ direction onto a sample with
surface at the $z=0$ plane, thickness $t$ in $z$ direction
and a large incidence area $A\rightarrow \infty$ in $(x,y)$ direction (see Fig. \ref{fig1}).

Instead of considering a three-dimensional diffusion
problem, in which the 
the material parameters are assumed to depend only on the $z$ direction,
as usually done \cite{chandrasekhar60,ishimaru78},
we consider the photon density $\bar U(z)$,
photon current $\bar j(z)$,
and source function $\bar \JJ(z)$, averaged
over the lateral ($x,y$) directions:
\ba
\bar U(z)&=&\frac{1}{A}\int_Adx\,dy\,U(\rr)\qquad
\bar j(z)=\frac{1}{A}\int\limits_Adx\,dy \,j_z(\rr)\nonumber\\
\bar \JJ(z)&=&\frac{1}{A}\int\limits_Adx\,dy \JJ(\rr)
\ea

It is evident that these quantities obey the following (quasi-) 
one-dimensional equations
\ba\label{diff2}
\frac{\partial}{\partial z}\bar j(z)&=&-\lambda_a\bar U(z)
+\bar \JJ(z)
\nonumber\\
\frac{\partial}{\partial z}\bar U(z)&=&
-\frac{1}{\widetilde D}\bar j(z)
\ea
which lead to the one-dimensional
diffusion equation
\be\label{diff3}
\lambda_a\bar U(z)=\widetilde D \frac{\partial^2}{\partial z^2}\bar U(z)
+\bar \JJ(z)
\ee
Defining now the incoming and outgoing currents as
\be\label{inout1}
I_\pm(z)=\frac{1}{2}\big[\bar U(z)\pm\bar j(z)\big]
\ee
we obtain from the diffusion equations (\ref{diff2}) 
the \mbox{Kubelka-Munk equations}
\ba\label{km1}
\bigg(\frac{\partial}{\partial z}+K\bigg)I_+(z)&=&-S\bigg(I_+(z)-I_-(z)\bigg)
+\bar \JJ(z)\nonumber\\
\bigg(\!-\frac{\partial}{\partial z}+K\bigg)I_-(z)&=&-S\bigg(I_-(z)-I_+(z)\bigg)
+\bar \JJ(z)
\nonumber\\
\ea
with
\ba\label{km2}
K&=&\lambda_a\nonumber\\
S&=&
\frac{1}{2}\bigg(\frac{1}{\widetilde D}-\lambda_a\bigg)
=\frac{3}{2}\lambda_t
\ea

Eq. (\ref{km2}) can also be written as
\be\label{km3}
\frac{1}{\widetilde D}=K+2S
\ee
\section{Derivation of reflectance and transmission coefficients}
Instead of solving Eqs. (\ref{km1}) we solve the
diffusion equation (\ref{diff3}).

The general solution of the homogeneous diffusion equation
(setting $\bar \JJ=0$ in Eq. (\ref{diff3}) ) is
\be\label{solution}
\bar U(z)=Ae^{\alpha z}
+Be^{-\alpha z}
\ee
whith the inverse diffusion length
\be\label{alpha}
\alpha=\sqrt{K/\widetilde D}=\sqrt{K(K+2S)}
\ee
From the solution (\ref{solution}) we get the in- and outgoing currents
\cite{wendlandt66,ciani05}
\be\label{inout}
I_\pm(z)=\frac{1}{2}\bigg(A(1\mp\beta)e^{\alpha z}+B(1\pm\beta)e^{-\alpha z}\bigg)
\ee
with 
\be
\beta=\widetilde D\alpha=\sqrt{K\widetilde D}=\sqrt{K/(K+2S)}
\ee
\subsection{Optically thick samples}
\subsubsection{No reflection at $z=0$}
The appropriate
boundary conditions corresponding to optically thick samples
without reflection at $z=0$ are
are
\be\label{boundaryinf}
I_+(0)=\bar U_0\qquad\qquad\bar I_+(\infty)=0
\ee
The second boundary condition implies $A=0$.
The in- and outgoing currents are therefore
\be\label{inout5}
I_\pm(z)=\frac{1}{2}B(1\pm\beta)e^{-\alpha z}
\ee
From the first boundary condition we obtain
\be
B=\bar U_0\frac{2}{1+\beta}
\ee
from wich we obtain the ingoing current at $z=0$
\be
I_-(0)=\bar U_0\frac{1-\beta}{1+\beta}
\ee
and hence the reflectivity
\be\label{rinf}
R_\infty=\frac{I_-(0)}{I_+(0)}=\frac{1-\beta}{1+\beta}
\ee

For the Kubelka-Munk function we obtain, using Eq. (\ref{km2})
\ba\label{km3}
\frac{S}{K}&=&\frac{1}{2}
\left[\left(\frac{1+R_\infty}{1-R_\infty}\right)^2-1\right]=
\frac{2R_\infty}{(1-R_\infty)^2}\nonumber\\
&&\nonumber\\
&=&\frac{3}{2}\frac{\ds \lambda_t}{\ds \lambda_a}
\ea
\subsubsection{Reflection at $z=0$}
The first boundary condition is now
\be
I_+(0)=U_0+R_0I_-(0)
\ee
where $R_0$ is the reflectivity at the $z=0$ boundary.
Inserting the expressions (\ref{inout5}) for $I_\pm(0)$
we get
\be
\frac{2}{1+\beta}I_+(0)=
B=
R_0\frac{2}{1+\beta}I_+(0)=
\frac{2}{1+\beta}U_0+R_0R_\infty B
\ee
from which follows
\be
B=\frac{1+\beta}{2}U_0\frac{1}{1-R_0R_\infty}
\ee
and hence
\be
R=\frac{1}{U_0}I_-(0)=\frac{R_\infty}{1-R_0R_\infty}
\ee
\subsection{Optically thin samples}
For optically thin samples with Reflectivity $R_1$ at the
back ($z=t$)
of the sample and Reflectivity $R_0$ at the front
($z=0$) of the sample
we have the boundary conditions
%\cite{wendlandt66,ciani05}
\be\label{boundary1}
I_+(0)=\bar U_0+R_0I_-(0)\qquad\qquad I_-(t)=R_1I_+(t)
\ee
Using the definition of $R_\infty$, Eq. (\ref{rinf}), 
we get from the boundary conditions a linear set of equations
for the coefficients $A$ and $B$,
which can be put into the form
\be
\left(
\begin{array}{cc}
	R_\infty-R_0&1-R_\infty R_0\\
	(1-R_\infty R_1)e^{\alpha t}&(R_\infty-R_1)e^{-\alpha t}
\end{array}
\right)
\left(
\begin{array}{c}A\\B
\end{array}
\right)
=
\left(
\begin{array}{c}
	\frac{2}{1+\beta}\bar U_0\\0
\end{array}
\right)
\ee
The determinant of the coefficient matrix is
\be
D=(R_\infty-R_0)(R_\infty-R_1)e^{-\alpha t}
-(1-R_\infty R_0)(1-R_\infty R_1)e^{\alpha t}
\ee
So we get from Kramer's rule
\be\label{a}
A=\frac{\bar U_0}{D}\frac{2}{1+\beta}e^{-\alpha t}(R_\infty-R_1)
\ee
\be\label{b}
B=-\frac{\bar U_0}{D}\frac{2}{1+\beta}e^{\alpha t}(1-R_1R_\infty)
\ee
We obtain for the currents at $z=0$ and at $z=t$:
\ba
I_-(0)&=&\frac{1+\beta}{2}\big[A+R_\infty B\big]\\
&=&\frac{\bar U_0}{D}\bigg[
	e^{-\alpha t}(R_\infty-R_1)-R_\infty e^{\alpha t}1-R_1R_\infty)
	\bigg]\nonumber
\ea
\ba
I_+(t)&=&\frac{1+\beta}{2}\big[R_\infty Ae^{\alpha t}+Be^{-\alpha t}\nonumber\\
&=&\frac{\bar U_0}{D}\bigg[
	R_\infty^2-1
	\bigg]\, ,
\ea
from which we ge the reflectivity $R$ 
\begin{widetext}
\be\label{r1}
R=\frac{I_-(0)}{\bar U_0}
=R_\infty\frac{
	e^{\alpha t}(1-R_\infty R_1)
	-e^{-\alpha t}(1-\frac{\ds R_1}{\ds R_\infty})
	}{
(1-R_\infty R_0)(1-R_\infty R_1)e^{\alpha t}
-(R_\infty-R_0)(R_\infty-R_1)e^{-\alpha t}
		}
\ee
and the transmittivity $T$
\be\label{t1}
T=\frac{I_+(t)}{\bar U_0}
=\frac{1-R_\infty^2
	}{
(1-R_\infty R_0)(1-R_\infty R_1)e^{\alpha t}
-(R_\infty-R_0)(R_\infty-R_1)e^{-\alpha t}
		}
\ee
Introducing the Kubelka-Munk parameters
\be
a=\frac{1}{2}\bigg(\frac{1}{R_\infty}+R_\infty\bigg)
\qquad
b=\alpha/S=\frac{1}{2}\bigg(\frac{1}{R_\infty}-R_\infty\bigg)
\ee
we get
\be
R=\frac{R_1b\cosh(\alpha t)R_1b+(1-R_1a)\sinh(\alpha t)
	}{
	b(1-R_0R_1)\cosh(\alpha t)
		+
		\big[
			a(1-R_0R_1)-R_0-R_1
			\big]
		\sinh(\alpha t)
		}
\ee
\be
T=\frac{b
	}{
	b(1-R_0R_1)\cosh(\alpha t)
		+
		\big[
			a(1-R_0R_1)-R_0-R_1
			\big]
		\sinh(\alpha t)
		}
\ee
\end{widetext}
If we set $R_0=0$, 
we get the formulas of
Kubelka (1948) \cite{kubelka48}
\be
R=\frac{
	1-R_1a+R_1b\coth(\alpha t)
	}{
		a-R_1+b\coth(\alpha t)
		}
\ee
and
\be
T=\frac{
	b
	}{
		b\cosh(\alpha t)+(a-R_1)\sinh(\alpha t)
		}
\ee
For $R_0=R_1=0$ we get the standard Kubelka-Munk formulas
\cite{kubelka48,wendlandt66,ciani05},
which do not contain the effect of internal reflections.
\be\label{r2}
R=\frac{\ds
	e^{\alpha t}
	+e^{-\alpha t}
	}{\ds
		e^{\alpha t}\frac{1}{R_\infty}
		-e^{-\alpha t}R_\infty
		}
=\frac{\sinh{\alpha t}}
{a\sinh{\alpha t}+b\cosh{\alpha t}}
\ee
\be\label{t2}
T=\frac{\ds\frac{1}{R_\infty}-R_\infty
	}{\ds
		e^{\alpha t}\frac{1}{R_\infty}
		-e^{-\alpha t}R_\infty
		}
=\frac{b}{
a\sinh{\alpha t}+b\cosh{\alpha t}}
\ee
Another interesting limit is that of very small $R_\infty$,
i.e. $R_\infty \rightarrow 0$:
\ba
R&=&
\frac{R_1e^{-\alpha t}}
{e^{\alpha t}-R_0R_1e^{-\alpha t}}\nonumber\\
&=&
\frac{R_1e^{-2\alpha t}}
{1-R_0R_1e^{-2\alpha t}}\nonumber\\
\ea
\ba
T&=&\frac{1}
{e^{\alpha t}-R_0R_1e^{-\alpha t}}\nonumber\\
&=&
\frac{e^{-\alpha t}}
{1-R_0R_1e^{-2\alpha t}}\nonumber\\
\ea
\section{Kubelka-Munk equations as one-dimensional radiative-transfer
equations}

We now want to demonstrate that the Kubelka-Munk
equations (\ref{km1}) are the proper radiadive-transfer equations
for the diffuse-reflection geometry depicted in \mbox{Fig. \ref{fig1}.}

We recall the three-dimensional
radiative transfer equations of the light intensity
in a turbid medium
\ba\label{radtransfer}
\big[\lambda_a+\sh\cdot\nabla\big]I(\rr,\sh)&=&
-\sum_{\sh'}q_{\sh\sh'}\bigg(I(\rr,\sh)-I(\rr,\sh')\bigg)\nonumber\\
&=&-\lambda_sI(\rr,\sh)+\sum_{\sh'}q_{\sh\sh'}I(\rr,\sh')
\ea
$I(\rr,\sh)$ is the distribution density of light rays passing through
$\rr$ with the direction $\sh=\kk/k$, where $\kk$ is the wave vector.
$q_{\sh\sh'}=|f(\sh,\sh')|^2$ is the phase function, i.e. the scattering
cross-section from $\sh$ to $\sh'$ with $f(\sh,\sh')$ being the corresponding
amplitude. $\sum_{\sh'}$ is an integral over the entire solid angle,
with the original direction $\sh$ being excluded.
The second line of Eq. (\ref{radtransfer}) is obtained from the sum rule
\be\label{sumrule}
\sum_{\sh'}q_{\sh\sh'}=
\sum_{\sh'}q_{\sh'\sh}
=\lambda_s
\ee

The three-dimensional
diffusion equations (\ref{diff1}) and (\ref{diff1a}) are
obtained from Eq. (\ref{radtransfer})
by ($i$) expanding the angle dependence of $I(\rr,\sh)$
and $q(\sh,sh')\approx q(\sh\cdot\sh')=q(\cos\gamma)$ 
in terms of Legendre polynomials and stop after the
1st term (P1 approximation) and then integrating $\sh$
over the total solid angle \cite{chandrasekhar60,ishimaru78}

The two terms of the three-dimensional
$I(\rr,\sh)$ in P1 approximation are
\cite{ishimaru78,porra97}
\be\label{diff3d}
I(\rr,\sh)=A_{3d}U(\rr)+B_{3d}\sh\cdot\jj(\rr)
\ee
with $U(\rr)=\sum\limits_\sh I(\rr,\sh)$, 
$\jj(\rr)=\sum\limits_\sh \sh I(\rr,\sh)$, 
and  $A_{3d}=1/\sum\limits_\sh=1/4\pi$ and
$B_{3d}=1/\sum\limits_\sh\sh\cdot\sh=3/4\pi$. 

The corresponding expression in one dimension is
\be
I(x,\sh)=A_{1d}U(x)+B_{1d}\sh\cdot\jj(x)
\ee
with $A_{1d}=B_{1d}=\sum\limits_\sh=
1/2$, which is just Eq. (\ref{inout1}).
Because we have shown in the beginning that
the diffusion equations (\ref{diff3}) are equivalent to the
Kubelka-Munk equations (\ref{km1}) we conclude that
the P1 approximation, and
hence the diffusion approximation
in one dimension is exact. This
has already been pointed out in Refs. \cite{porra97,goldstein51}.

So we can state that the Kubelka-Munk equations (\ref{km1}) are
($i$) identical to the three-dimensional diffusion equation,
averaged over the lateral dimensions, and ($ii$) are the proper
radiative-transfer equations for the one-dimensional 
diffuse-reflection problem.

\section{Discussion}
We now turn to the previously published derivations of the
Kubelka-Munk equations from the radiative-transfer equations
(\ref{radtransfer}) 
\cite{ishimaru78,edstroem03,edstroem04,sandoval14}.

All these authors start from the three-dimensional radiative transfer
equation, in which the material parameters are assumed only to depend
on the coordinate $z$ but the light rays still retain their
three-dimensional orientations, parametrized
by $\mu=\sh\cdot{\bf e}_x=\cos\theta$:
\be\label{radtransfer1}
\big[\lambda_a+\mu\frac{\partial}{\partial x}\big]I(x,\mu)
=-\lambda_sI(x,\mu)+\frac{1}{2}\int_{-1}^1d\mu q(\mu,\mu')I(x,\mu')
\ee
They then identify the contributions to $I(x,\mu)$ in positive and
negative directions as
\be
\widetilde I_{\pm}(x,\mu)=\theta(\pm x)I(x,\mu)
\ee
where $\theta(x)$ is the Heaviside step function. The in- and outgoing
currents are defined as
\be
\widetilde I_\pm(x)=\frac{1}{2}\int_{-1}^1d\mu I_\pm(x,\mu)
\ee
Then the diffusion approximation is done, which, from Eq. (\ref{diff3d})
gives \cite{ishimaru78,gate71}
\ba
\widetilde I_\pm(x)&=&\frac{1}{4}U(x)\pm\frac{1}{2}j_x(x)\nonumber\\
&=&\frac{1}{4}U(x)\mp\frac{\widetilde D}{2}\frac{\partial}{\partial x}U(x)
\ea
However, this equation violates the requirement
$I_+(x)+I_-(x)=\bar U(x)$. Obviously
this is the reason, why in Refs. 
\cite{ishimaru78,edstroem03,edstroem04,sandoval14} $K$ is identified
with $2\lambda_a$, instead of the correct result $K=\lambda_a$.

Diffusive reflection with collimated, instead of uniform, illumination
may be readily treated in the three-dimensional
diffusion (P1) approximation, which, however,
is not the subject of the present treatment.

\vspace*{.4cm}

\section{Conclusion}
We have shown that the Kubelka-Munk equations are identical
to the one-dimensional diffusion equation, which is obtained by
averaging the three-dimensional diffusion equation with respect to
the lateral directions. We obtain as Kubelka-Munk parameters
$K=\lambda_a$ (absorptive inverse scattering length) and
$S=\frac{3}{2}\lambda_t=
\frac{3}{2}\lambda_s(1-\langle\cos\gamma\rangle)$, where
$\lambda_t$ and $\lambda_s$ are the transport and scattering
inverse scattering lengths, and $\langle\cos\gamma\rangle$
is the anisotropy parameter.
Using the 1d diffusion equation we have derived formulas
for the diffuse reflection and transmission, which includes
possible internal reflections. In the absence of internal
reflections these expressions reduce to those given
by Kubelka and Munk.
We have demonstrated that the Kubelka-Munk equations are the
appropriate radiative transfer equations for the
reflection problem
with plane-wave incidence (uniform illumination). 

\end{document}